\documentclass[
reprint,
aps,
prd,
superscriptaddress,
nofootinbib,
longbibliography
]{revtex4-2}

\usepackage{amsmath,amssymb,amsfonts}
\usepackage[T1]{fontenc}
\usepackage[utf8]{inputenc}
\usepackage{hyperref}
\usepackage{graphicx}
\usepackage{tikz}
\usetikzlibrary{decorations.pathmorphing}

\newcommand{\Bsbx}{\mathcal{B}_{\rm sbx}}
\newcommand{\Bsprx}{\mathcal{B}_{\rm sprx}}
\newcommand{\Ext}{\mathcal{E}}
\newcommand{\Cf}{\mathcal{C}_{\rm f}}
\newcommand{\Crel}{\mathcal{C}_{\rm rel}}
\newcommand{\GN}{G_N}
\newcommand{\Om}{\Omega_{D-2}}

\begin{document}

\title{Computational Cosmic Censorship}

\author{Fuat Berkin Altunkaynak}
\affiliation{H\"useyin Avni S\"ozen Anatolian High School, \"Usk\"udar 34662, Istanbul, Turkey}
\email{fuatberkin34@gmail.com}

\begin{abstract}
We identify a two-gap complexity structure underlying a holographic obstruction to reaching extremality and to the black-hole-mediated route toward weak cosmic censorship violation. The complexity of formation of extremal AdS black holes diverges logarithmically, while it is finite for every subextremal member of the stationary family. We compute the on-shell Wheeler--DeWitt action of overcharged Reissner--Nordstr\"om--AdS in general dimension $D\geq4$. In the minimal prescription, defined by omission of the Maxwell boundary term, the individually divergent Maxwell bulk, on-shell Einstein--Hilbert, and Gibbons--Hawking--York terms cancel exactly. We also evaluate the null--timelike joints and null-boundary counterterms explicitly and show that their combined lower-endpoint contribution vanishes. Adding the Maxwell boundary term produces only a prescription-dependent, regulator-local power-law divergence; no prescription produces a logarithmic divergence, and the corresponding vacuum-subtracted complexity=volume complexity of formation is finite. In the complexity=action and complexity=volume prescriptions, the reverse triangle inequality for complexity distance then implies divergent relative complexity across each adjacent gap. Under a finite-complexity-growth assumption, this gives a conditional obstruction to traversing the stationary path from subextremal black holes through extremality to the superextremal naked-singularity sector.
\end{abstract}
\maketitle

\section{Introduction}
The weak cosmic censorship conjecture, originally proposed by Penrose~\cite{Penrose1969}, posits that
singularities arising in gravitational collapse are generically hidden behind event horizons and are
therefore inaccessible to distant observers. Despite strong evidence in many settings, no general proof
is known, and explicit violations can occur under special conditions. This raises a fundamental question:
what principle, if any, enforces the exclusion of naked singularities from the physically realizable
sector?

Conventional formulations of cosmic censorship are geometric, relying on causal structure and global
properties of spacetime. However, they do not directly address whether such geometries are operationally
accessible. In particular, even if a spacetime containing a naked singularity exists as a classical
solution, it is not clear whether it can be formed through any finite physical process. This suggests
that cosmic censorship may admit an alternative formulation in which the relevant constraint is not
purely geometric, but dynamical or computational.

A natural framework for exploring such questions is provided by the AdS/CFT
correspondence~\cite{Maldacena1998}, which relates quantum gravity in $D$-dimensional asymptotically
anti-de Sitter spacetimes to conformal field theories in $D-1$ dimensions. In this setting, bulk
geometries correspond to boundary states that can, in principle, be prepared by unitary evolution.
This motivates characterizing the accessibility of a spacetime in terms of the computational resources
required to prepare its dual state. In particular, the complexity=action (CA)
proposal~\cite{Brown2016PRL,Brown2016PRD} identifies the quantum complexity of a boundary state with
the gravitational action of the associated Wheeler--DeWitt (WdW) patch, while the complexity=volume
(CV) proposal~\cite{SusskindStanford} identifies it with the volume of an extremal bulk slice. The
complexity of formation is the corresponding vacuum-subtracted cost of preparing the state relative
to the AdS vacuum~\cite{ChapmanFormation2017}. We denote it by $\Cf$, with $\Cf^{\rm A}$ and
$\Cf^{\rm V}$ when the CA and CV prescriptions must be distinguished. Reviews of circuit and
holographic complexity include~\cite{ChapmanPolicastro2022}.

The operational quantity needed below is the relative complexity between two boundary states,
\begin{equation}
\Crel(X,Y)\equiv\min_{U:\,U|X\rangle=|Y\rangle}\mathcal{C}(U).
\label{eq:relative-complexity}
\end{equation}
Relative complexity measures the least circuit cost of making the transition~\cite{BrownSusskind2018}. Nielsen's
original geometric construction equips the unitary group with a penalized geometry and defines the
operator cost $\mathcal{C}(U)$ as a shortest-path length~\cite{Nielsen2006}. Equation~\eqref{eq:relative-complexity}
is instead the induced state complexity: one further minimizes that unitary cost over all endpoint
unitaries mapping $|X\rangle$ to $|Y\rangle$, equivalently passing to the quotient by the stabilizer of
the state~\cite{Auzzi2021}. For the reversible metric cost assumed in the distance argument, this construction
induces a state-space distance $d(X,Y)=\Crel(X,Y)$. If $\mathcal{C}_{R}(X)=d(R,X)$ denotes the complexity of
$X$ relative to a common reference state $R$, the reverse triangle inequality gives
\begin{equation}
\Crel(X,Y)=d(X,Y)\geq
\left|\mathcal{C}_{R}(X)-\mathcal{C}_{R}(Y)\right|.
\label{eq:reverse-triangle}
\end{equation}
Consequently, a divergent difference of formation complexities relative to the same vacuum reference
is sufficient, without identifying that difference with the exact optimal circuit, to establish
divergent transition complexity.

A second ingredient is that complexity should not change arbitrarily fast under physical time
evolution. Rigorous results in random and local random circuit models show linear complexity growth
for long times in settings designed to capture generic chaotic dynamics
\cite{Haferkamp2021,Brandao2021}. Throughout this paper we assume that physically realizable boundary
evolutions have finite complexity growth rate. This assumption is weaker than imposing a Lloyd-type
upper bound~\cite{Lloyd2000}; relaxing it would affect CA- and CV-based arguments generally, not the
present mechanism specifically. Together with Eq.~\eqref{eq:reverse-triangle}, it implies that states
at divergent relative complexity cannot be connected in finite boundary time.

Recent developments suggest that such an operational perspective may be useful. Cryptographic
censorship relates pseudorandom boundary evolution to the existence of a bulk event
horizon~\cite{Engelhardt2025}, while complexity has also been used as a probe of strong cosmic
censorship~\cite{Alishahiha2022} and in complexity-based analyses of black-hole
singularities~\cite{Mohan2025}. Studies of non-isometric bulk-to-boundary maps further indicate that
not all semiclassical bulk configurations correspond to efficiently realizable boundary states, with
pathological configurations suppressed by complexity~\cite{Akers2024}. The present work is
complementary: it studies the complexity geometry of explicit stationary spacetime sectors rather
than pseudorandom evolution or a general singularity theorem.

The key observation is that both the third law of black-hole mechanics and weak cosmic censorship, in
their overcharging form, concern the crossing of the extremality bound. The classical third law was
formulated as an unattainability statement in~\cite{Bardeen1973,Israel1986}. Its unrestricted
form is not universal: finite-time dynamical formation of extremal horizons has been constructed in
Einstein--Maxwell--charged-scalar systems, including extensions with nonzero cosmological constant
\cite{KehleUnger2025,Gadioux2026}, while conditional AdS unattainability results survive under
additional supersymmetric energy--charge assumptions~\cite{McSharryReall2025}. The result below
therefore does not assume a universal classical third law. It concerns a finite-complexity-growth path
through the stationary holographic family.

We denote a subextremal black hole by $\Bsbx$, the extremal solution by $\Ext$, and the superextremal
overcharged geometry by $\Bsprx$. Along the stationary family there are two adjacent gaps,
\begin{equation}
\Bsbx \longrightarrow \Ext \longrightarrow \Bsprx .
\end{equation}
For the first gap, the complexity of formation of charged or rotating AdS black holes diverges
logarithmically at extremality~\cite{Carmi2017,Balushi2020,Bernamonti2021}, whereas $\Cf$ is finite for
any fixed subextremal member. Equation~\eqref{eq:reverse-triangle} will then imply a divergent relative
complexity. For the second gap, the relevant question is whether the naked-singularity side contains a
matching logarithmic divergence capable of matching the extremal divergence. The central result of this paper
is that it does not in the CA prescription or in CV.

We compute the WdW action of overcharged RN--AdS in general dimension. The Einstein--Hilbert bulk term is finite for $D=4$, but diverges for $D>4$: on shell,
\begin{equation}
R-2\Lambda=\frac{4\Lambda}{D-2}+\frac{D-4}{D-2}F^2,
\label{eq:trace-intro}
\end{equation}
so the Einstein--Hilbert term inherits the same $\epsilon^{-(D-3)}$ near-singularity divergence as the
Maxwell term. The correct cancellation is therefore a three-way cancellation. In units defined below,
the Maxwell bulk, on-shell Einstein--Hilbert, and Gibbons--Hawking--York (GHY) terms contribute
$+(D-2)$, $-(D-4)$, and $-2$, whose sum vanishes identically. Prescription dependence enters only when
the Maxwell boundary term is added; its residual divergence is a pure power law and never a logarithm.
The near-origin CV contribution also vanishes, and the vacuum-subtracted CV complexity of formation is
finite. Thus, in CA and CV,
Eq.~\eqref{eq:reverse-triangle} gives a divergent relative complexity between $\Ext$ and $\Bsprx$.
Nonminimal Maxwell boundary prescriptions add a regulator-local power law but no logarithm. The absence of a matching logarithmic divergence, and hence the operational distance argument, is unaffected by this prescription dependence. Under finite complexity
growth, a quasi-stationary Wald-type overcharging path cannot traverse the stationary extremal surface
in finite boundary time. The statement is not a proof of full WCCC; it is a conditional holographic
obstruction to the black-hole-mediated route from $\Bsbx$ to $\Bsprx$.

\begin{figure*}[t]
    \centering
    \begin{tikzpicture}[
      x=0.85cm,
      y=0.85cm,
      line cap=round,
      line join=round
    ]

    \definecolor{patchblue}{RGB}{143,167,216}
    \definecolor{edgeblue}{RGB}{38,67,103}

    \begin{scope}[shift={(0,0)}]

      \coordinate (L) at (0,3.35);
      \coordinate (R) at (5.35,3.35);
      \coordinate (T) at (2.675,6.05);
      \coordinate (B) at (2.675,0.65);

      \draw[
        black,
        line width=0.65pt
      ] (L) -- (0,-1.25);

      \draw[
        black,
        line width=0.65pt
      ] (R) -- (5.35,-1.25);

      \draw[
        black,
        line width=0.85pt,
        decorate,
        decoration={
          zigzag,
          segment length=5.2pt,
          amplitude=2.7pt
        }
      ] (L) -- (0,7.45);

      \draw[
        black,
        line width=0.85pt,
        decorate,
        decoration={
          zigzag,
          segment length=5.2pt,
          amplitude=2.7pt
        }
      ] (R) -- (5.35,7.45);

      \draw[
        dash pattern=on 5pt off 5pt,
        black,
        line width=0.55pt
      ] (3.90,7.25) -- (L);

      \draw[
        dash pattern=on 5pt off 5pt,
        black,
        line width=0.55pt
      ] (1.45,7.25) -- (R);

      \draw[
        dash pattern=on 5pt off 5pt,
        black,
        line width=0.55pt
      ] (0.075,3.25) -- (B) -- (4.575,-1.25);

      \draw[
        dash pattern=on 5pt off 5pt,
        black,
        line width=0.55pt
      ] (5.275,3.25) -- (B) -- (0.775,-1.25);

      \filldraw[
        fill=patchblue,
        fill opacity=0.78,
        draw=edgeblue,
        line width=1.05pt
      ]
        (0,2.8) -- (2.675,5.5) -- (5.35,2.8) -- (2.675,0.15) -- cycle;

      \draw[
        dash pattern=on 5pt off 5pt,
        black,
        line width=0.55pt
      ] (0.075,3.25) -- (B);

      \draw[
        dash pattern=on 5pt off 5pt,
        black,
        line width=0.55pt
      ] (5.275,3.25) -- (B);

      \node[
        anchor=east
      ] at (0,2.8) {$t_L$};

      \node[
        anchor=west
      ] at (5.35,2.8) {$t_R$};

      \node[
        anchor=east
      ] at (-0.13,7.16) {$r=0$};

      \node[
        anchor=west
      ] at (5.48,7.16) {$r=0$};

      \node[
        anchor=east
      ] at (-0.13,0.98) {$r=\infty$};

      \node[
        anchor=west
      ] at (5.48,0.98) {$r=\infty$};

      \node[
        rotate=45,
        fill=white,
        inner sep=1pt
      ] at (1.10,4.95) {$r=r_{-}$};

      \node[
        rotate=-45,
        fill=white,
        inner sep=1pt
      ] at (4.25,4.95) {$r=r_{-}$};

      \node[
        rotate=-45,
        inner sep=1pt
      ] at (1.30,2.25) {$r=r_{+}$};

      \node[
        rotate=45,
        inner sep=1pt
      ] at (4.0,2.25) {$r=r_{+}$};

      \node[
        rotate=45,
        fill=white,
        inner sep=1pt
      ] at (1.18,-0.52) {$r=r_{+}$};

      \node[
        rotate=-45,
        fill=white,
        inner sep=1pt
      ] at (4.18,-0.52) {$r=r_{+}$};

      \node at (2.675,-1.50) {\large (a)};

    \end{scope}

    \begin{scope}[shift={(8.10,0)}]

      \coordinate (LT) at (0.05,5.72);
      \coordinate (LB) at (0.05,0.10);
      \coordinate (A)  at (4.32,2.95);

      \filldraw[
        fill=patchblue,
        fill opacity=0.78,
        draw=edgeblue,
        line width=1.05pt
      ]
        (LT) -- (A) -- (LB) -- cycle;

      \draw[
        black,
        line width=0.85pt,
        decorate,
        decoration={
          zigzag,
          segment length=5.2pt,
          amplitude=2.7pt
        }
      ] (0,-1.25) -- (0,7.45);

      \node[
        anchor=west
      ] at (4.42,2.95) {$t_b$};
      \draw[
        black,
        line width=0.70pt
      ] (4.32,-1.25) -- (4.32,7.45);

      \node at (0.05,-1.40) {$r=0$};

      \node at (4.32,-1.40) {$r=\infty$};

      \node at (2.18,-1.50) {\large (b)};

    \end{scope}

    \end{tikzpicture}
    \caption{(a) Eternal Reissner--Nordström--AdS black hole with two asymptotic boundaries at $r=\infty$ (left and right). Dashed lines mark the outer ($r=r_{+}$) and inner ($r=r_{-}$) horizons, while zig-zag lines indicate timelike singularities at $r=0$. The Wheeler–DeWitt patch does not extend to the singularity. (b) Over-charged (superextremal) Reissner--Nordström--AdS spacetime in which the horizons have disappeared and the timelike singularity at $r=0$ is exposed (naked). Only a single AdS boundary at $r=\infty$ remains (vertical line). The shaded triangular Wheeler–DeWitt patch is anchored at $t_b$ and bounded by null rays that end on the singularity.}
    \label{fig:rnads}
\end{figure*}
\section{Complexity=Action Computation for Overcharged RN--AdS}
See Fig.~\ref{fig:rnads} for Penrose diagrams and corresponding Wheeler--DeWitt patches.
The $D$-dimensional metric is
\begin{equation}
ds^2=-f(r)\,dt^2+\frac{dr^2}{f(r)}+r^2 d\Omega^2_{D-2},
\label{eq:metric}
\end{equation}
with
\begin{equation}
f(r)=1+\frac{r^2}{L^2}-\frac{\mu}{r^{D-3}}+\frac{q^2}{r^{2D-6}},
\label{eq:f}
\end{equation}
solving Einstein--Maxwell theory with cosmological constant
$\Lambda=-(D-1)(D-2)/(2L^2)$. We use the normalization
\begin{equation}
I_{\rm bulk}=\frac{1}{16\pi \GN}\int d^D x\sqrt{-g}\,
\left(R-2\Lambda-F_{\mu\nu}F^{\mu\nu}\right),
\label{eq:bulk-action}
\end{equation}
with gauge field
\begin{equation}
A_t(r)=\Phi-\sqrt{\frac{D-2}{2(D-3)}}\frac{q}{r^{D-3}},\qquad
F_{tr}=-c_D\frac{q}{r^{D-2}},
\label{eq:gaugefield}
\end{equation}
where $c_D^2=(D-2)(D-3)/2$. For fixed $q$ and $L$ there is a critical mass parameter
$\mu_{\rm ext}(q,L)$ at which $f$ has a double root. In the overcharged regime
$\mu<\mu_{\rm ext}$, $f(r)$ has no positive root and the timelike curvature singularity at $r=0$
enters the WdW patch directly.

The full action is
\begin{equation}
I_{\rm WdW}=I_{\rm bulk}+I_{\rm GHY}+I_{\rm null}+I_{\rm ct}+I_{\rm joints}+I_\gamma,
\label{eq:fullaction}
\end{equation}
where the optional Maxwell boundary term is
\begin{equation}
I_\gamma=\frac{\gamma}{4\pi \GN}\int_{\partial M}d\Sigma_\mu\,F^{\mu\nu}A_\nu.
\label{eq:maxwell-boundary}
\end{equation}
Here $\gamma=0$ is the minimal Dirichlet prescription, while $\gamma=1$ gives the fixed-charge
prescription~\cite{Goto2019}. The WdW patch anchored at $t=0$ is bounded by radial null surfaces
satisfying
\begin{equation}
\frac{dt}{dr}=\pm \frac{1}{f(r)},
\label{eq:null}
\end{equation}
with time width
\begin{equation}
\Delta t(r)=2\int_r^{r_{\rm max}}\frac{dx}{f(x)}.
\label{eq:deltat}
\end{equation}
Near the singularity,
\begin{equation}
f(r)\sim \frac{q^2}{r^{2D-6}},\qquad
\Delta t(r)=\Delta t_0+O(r^{2D-5}),
\label{eq:fnear}
\end{equation}
so the patch has finite time extent at $r=0$. Any singular contribution is therefore local in the
near-origin action.

The trace of the Einstein equations gives
\begin{align}
R&=\frac{2D\Lambda}{D-2}+\frac{D-4}{D-2}F^2,\notag\\
R-2\Lambda-F^2&=\frac{4\Lambda}{D-2}-\frac{2}{D-2}F^2.
\label{eq:trace}
\end{align}
with
\begin{equation}
F^2=-\frac{2c_D^2 q^2}{r^{2D-4}}.
\label{eq:F2}
\end{equation}
Using $\sqrt{-g}=\Omega_{D-2}\,r^{D-2}$, the on-shell bulk action becomes,
\begin{equation}
\begin{split}
I_{\rm bulk}
&=\frac{\Om}{16\pi\GN}\int_\epsilon^{r_{\rm max}}dr\,\Delta t(r)\\
&\quad\times\left[\frac{4\Lambda r^{D-2}}{D-2}+\frac{2(D-3)q^2}{r^{D-2}}\right].
\end{split}
\label{eq:bulk-onshell}
\end{equation}
The cosmological term is integrable at $r=0$. The charge term gives
\begin{equation}
I_{\rm bulk}^{\rm div}=+\frac{\Om\Delta t_0 q^2}{8\pi\GN}\frac{1}{\epsilon^{D-3}}.
\label{eq:bulk-div}
\end{equation}
It is useful to introduce
\begin{equation}
u\equiv \frac{\Om\Delta t_0 q^2}{16\pi\GN}\frac{1}{\epsilon^{D-3}}.
\label{eq:u}
\end{equation}
Then the Maxwell part $-F^2/(16\pi\GN)$ contributes $+(D-2)u$, while the on-shell Einstein--Hilbert
part $(R-2\Lambda)/(16\pi\GN)$ contributes $-(D-4)u$. Thus the Einstein--Hilbert term is finite only in
$D=4$; for $D>4$ it is an essential part of the cancellation.

We regulate the naked singularity by a timelike surface $r=\epsilon$. With outward-pointing
(decreasing-$r$) unit normal,
\begin{equation}
I_{\rm GHY}=\frac{1}{8\pi\GN}\int d^{D-1}x\sqrt{|h|}\,K,
\label{eq:ghy}
\end{equation}
where
\begin{equation}
\sqrt{|h|}=r^{D-2}\sqrt{f(r)},\qquad
K=-\frac{f'(r)}{2\sqrt{f(r)}}-\frac{D-2}{r}\sqrt{f(r)}.
\label{eq:hK}
\end{equation}
Therefore
\begin{equation}
\begin{split}
I_{\rm GHY}(r=\epsilon)
&=\frac{\Om}{8\pi\GN}\Delta t(\epsilon)
\Big[-\frac{1}{2}\epsilon^{D-2}f'(\epsilon) \\
&\hspace{2.2cm} -(D-2)\epsilon^{D-3}f(\epsilon)\Big].
\end{split}
\label{eq:ghy-combined}
\end{equation}
Substituting Eq.~\eqref{eq:f} gives
\begin{equation}
\begin{split}
I_{\rm GHY}(\epsilon)
&=\frac{\Om\Delta t(\epsilon)}{8\pi\GN}
\Big[-\frac{q^2}{\epsilon^{D-3}}+\frac{D-1}{2}\mu \\
&\quad -(D-2)\epsilon^{D-3}
-\frac{D-1}{L^2}\epsilon^{D-1}\Big].
\end{split}
\label{eq:ghy-expanded}
\end{equation}
So
\begin{equation}
I_{\rm GHY}^{\rm div}=-\frac{\Om\Delta t_0 q^2}{8\pi\GN}\frac{1}{\epsilon^{D-3}}=-2u.
\label{eq:ghydiv}
\end{equation}

We next evaluate the null terms rather than infer their scaling. Let $n=D-2$ and choose the two null
normal one-forms
\begin{equation}
k^{(\pm)}_\mu dx^\mu=\alpha_\pm\left(\pm dt+\frac{dr}{f(r)}\right),
\qquad \alpha_\pm>0 .
\label{eq:null-normals}
\end{equation}
Because these normals are gradients of null coordinates with constant normalization, their generators
are affinely parametrized, so the null-boundary term proportional to the nonaffinity vanishes,
\begin{equation}
I_{\rm null}=0.
\label{eq:null-zero}
\end{equation}
For the timelike regulator, $n_\mu dx^\mu=-dr/\sqrt{f}$, and hence
$|k^{(\pm)}\!\cdot n|=\alpha_\pm/\sqrt{f(\epsilon)}$. The future and past null boundaries
intersect the regulator $r=\epsilon$ in two null--timelike joints. With the joint conventions of~\cite{Lehner2016}, their combined contribution is
\begin{align}
I_{\rm joint}^{(\epsilon)}
&=-\frac{\Om\epsilon^n}{8\pi\GN}
 \left[\ln\frac{\alpha_+}{\sqrt{f(\epsilon)}}+
       \ln\frac{\alpha_-}{\sqrt{f(\epsilon)}}\right]\notag\\
&=-\frac{\Om\epsilon^n}{8\pi\GN}
 \ln\frac{\alpha_+\alpha_-}{f(\epsilon)} .
\label{eq:joint-explicit}
\end{align}

The reparametrization-invariant counterterm on each null boundary is
\begin{equation}
I_{\rm ct}=\frac{1}{8\pi\GN}\sum_{\pm}
\int_{\mathcal{N}_\pm}d\lambda\,d^n\theta\,
\sqrt{\gamma}\,\Theta\,
\ln\left(\frac{\ell_{\rm ct}|\Theta|}{n}\right),
\label{eq:null-counterterm}
\end{equation}
where $\ell_{\rm ct}$ is an arbitrary length scale
\cite{Lehner2016,ReynoldsRoss2017}. Since
$\sqrt{\gamma}=r^n\sqrt{\omega}$, $|dr/d\lambda|=\alpha_\pm$, and
$|\Theta|=n\alpha_\pm/r$, direct integration gives the lower-endpoint contribution
\begin{align}
I_{\rm ct}^{(\epsilon)}
&=\frac{\Om\epsilon^n}{8\pi\GN}
\left[
\ln\frac{\ell_{\rm ct}^2\alpha_+\alpha_-}{\epsilon^2}
+\frac{2}{n}
\right].
\label{eq:ct-explicit}
\end{align}
The signs in Eqs.~\eqref{eq:joint-explicit} and \eqref{eq:ct-explicit} follow from orienting
the null generators from the asymptotic cutoff toward the inner regulator. Reversing all boundary
orientations changes both endpoint signs but not the conclusion. Their sum is independent of the
arbitrary affine normalizations:
\begin{align}
I_{\rm joint}^{(\epsilon)}+I_{\rm ct}^{(\epsilon)}
&=\frac{\Om\epsilon^n}{8\pi\GN}
\left[
\ln\frac{\ell_{\rm ct}^2 f(\epsilon)}{\epsilon^2}
+\frac{2}{n}
\right]\notag\\
&=\frac{\Om\epsilon^{D-2}}{8\pi\GN}
\Big[
-2(D-2)\ln\epsilon+\ln(\ell_{\rm ct}^2q^2)
\notag\\
&\hspace{2.0cm}
+\frac{2}{D-2}+O(\epsilon^{D-3})
\Big]\notag\\
&=O(\epsilon^{D-2}\ln\epsilon)\longrightarrow0 .
\label{eq:joint-ct-sum}
\end{align}
Thus the logarithms carried by the joint and counterterm are suppressed by the vanishing
area of the regulator sphere and cannot reproduce the extremal logarithm. The upper-endpoint pieces
belong to the standard AdS UV structure and cancel in the vacuum-subtracted complexity of formation
\cite{ChapmanFormation2017,ReynoldsRoss2017}.

Collecting the divergent terms, the minimal prescription gives
\begin{equation}
I_{\rm div}^{\gamma=0}=\big[(D-2)-(D-4)-2\big]u=0.
\label{eq:exact-cancel}
\end{equation}
Thus the overcharged RN--AdS WdW action has no near-singularity divergence in the minimal
prescription. After the standard vacuum subtraction of the AdS UV divergences, the corresponding CA
complexity of formation is finite. The cancellation is structural: the divergent on-shell bulk integrand is a
total radial derivative, whose lower-endpoint value is precisely cancelled by the GHY term, as shown in
Appendix~\ref{app:total-derivative}.

When Eq.~\eqref{eq:maxwell-boundary} is included, its regulator contribution gives
\begin{equation}
I_\gamma^{\rm div}=-2\gamma(D-2)u,
\label{eq:gamma-div}
\end{equation}
with a gauge-independent divergent coefficient. Hence the net near-singularity action is
\begin{equation}
\begin{split}
I_{\rm WdW}^{\rm ns}
&=K_D(\gamma)\frac{\Om\Delta t_0 q^2}{16\pi\GN}\frac{1}{\epsilon^{D-3}},\\
K_D(\gamma)&=-2\gamma(D-2).
\end{split}
\label{eq:fulldivcombined}
\end{equation}
The residual term vanishes only at $\gamma=0$; for $\gamma\ne0$ it is a prescription-dependent,
regulator-local pure power law. In no prescription is there a near-singularity logarithm. The
operational distance argument below therefore uses the finite minimal-CA complexity of formation (and
independently the finite CV complexity of formation). For $\gamma\ne0$, the robust conclusion is
narrower: the added local term supplies no
universal logarithm that could be identified with the extremal throat divergence. The near-singularity contributions to the WdW action of overcharged RN--AdS are summarized in Table~\ref{tab:divergences}.

\begin{table*}[t]
\caption{Near-singularity contributions to the WdW action of overcharged RN--AdS as $\epsilon\to0$, in units of $u=\Om\Delta t_0 q^2/(16\pi\GN\epsilon^{D-3})$. The joint and counterterm entries refer to their explicit lower-endpoint sum in Eq.~\eqref{eq:joint-ct-sum}. No term produces an unsuppressed logarithmic divergence.}
\label{tab:divergences}
\begin{ruledtabular}
\begin{tabular}{lcc}
Contribution & Scaling & Coefficient \\
\hline
Maxwell bulk & $\epsilon^{-(D-3)}$ & $+(D-2)$ \\
Einstein--Hilbert, on shell & $\epsilon^{-(D-3)}$ & $-(D-4)$ \\
GHY at $r=\epsilon$ & $\epsilon^{-(D-3)}$ & $-2$ \\
Null boundaries & -- & $0$ \\
Joint $+$ null counterterm & $\epsilon^{D-2}\ln\epsilon$ & $0$ \\
Maxwell boundary term & $\epsilon^{-(D-3)}$ & $-2\gamma(D-2)$ \\
Total & $\epsilon^{-(D-3)}$ & $-2\gamma(D-2)$
\end{tabular}
\end{ruledtabular}
\end{table*}

Within CA,
\begin{equation}
\mathcal{C}_{\rm A}=\frac{I_{\rm WdW}}{\pi\hbar}.
\label{eq:CA}
\end{equation}
Therefore, after the standard vacuum subtraction, $\Cf^{\rm A}(\Bsprx)$ is finite in the minimal
prescription. Nonminimal Maxwell boundary prescriptions add the local power law in
Eq.~\eqref{eq:fulldivcombined}, but none generates a near-singularity logarithm.

\section{Complexity=volume Computation}
The same conclusion holds for the CV complexity of formation. Since the superextremal geometry has
$f(r)>0$ for $r>0$, the maximal slice anchored at a boundary time $t_b$ is the static slice $t=t_b$.
For a spherically symmetric graph $t=t(r)$, the radial volume density is proportional to
$r^{D-2}\sqrt{f^{-1}-f[t'(r)]^2}$ and is pointwise maximized by $t'(r)=0$.
\begin{equation}
V(\epsilon,r_{\rm max})=\Om\int_\epsilon^{r_{\rm max}}dr\,\frac{r^{D-2}}{\sqrt{f(r)}}.
\label{eq:cv-volume-rnads}
\end{equation}
Near $r=0$, $f(r)\sim q^2/r^{2D-6}$, hence
\begin{equation}
V(0,\epsilon)\sim \frac{\Om}{q}\int_0^\epsilon dr\,r^{2D-5}=O(\epsilon^{2D-4}).
\label{eq:cv-rnads-origin}
\end{equation}
Thus the naked singularity gives no logarithmic or power-law divergence. The fact that the slice reaches
$r=0$ is not inconsistent with maximality: maximal slices avoid spacelike black-hole singularities because
of the black-hole causal structure, not as part of the definition of CV. Here there is no horizon and the
singularity is timelike, so the variational problem is naturally posed first on the regulated spacetime
$r\geq\epsilon$, where the slice ends on the timelike regulator. Equation~\eqref{eq:cv-rnads-origin}
shows that this endpoint contributes vanishing volume as $\epsilon\to0$. The slice is therefore
well behaved for the regulated CV calculation, although it should not be interpreted as a smooth ``nice
slice'' through a geodesically complete geometry; related CV analyses of timelike singularities appear in~\cite{Katoch2023}.

The only remaining divergence is the standard AdS UV volume divergence, removed in the complexity of
formation~\cite{ChapmanFormation2017,ReynoldsRoss2017}. Hence $\Cf^{\rm V}(\Bsprx)$ is finite and
carries no logarithmic divergence. On the extremal side, by contrast, the CV logarithm comes from the
near-horizon throat: at extremality $f$ has a double root, so $1/\sqrt{f}\sim1/(r-r_e)$ and the exterior
static volume contains a logarithmic throat integral. The superextremal origin crushes the integrand rather
than enhancing it, so the analytic structure on the naked-singularity side cannot match the extremal
throat logarithm.

There is also no analogue of the usual late-time black-hole growth in this stationary one-boundary
geometry. A WdW patch anchored at $t_b$ and the maximal slice $t=t_b$ are obtained from their $t_b=0$
counterparts by the static Killing translation. With a time-translation-invariant regulator prescription at
the timelike singularity, both vacuum-subtracted quantities are therefore independent of $t_b$,
\begin{equation}
\frac{d\Cf^{\rm A}}{dt_b}=\frac{d\Cf^{\rm V}}{dt_b}=0.
\label{eq:zero-time-growth}
\end{equation}
Thus linear growth is not expected here: the mechanism responsible for it in eternal black holes, namely
the growth of the region behind a horizon, is absent. Time-dependent perturbations or time-dependent
boundary conditions at the timelike singularity could change this conclusion, but they lie outside the
stationary calculation. In particular, the vanishing rates satisfy the finite-growth assumption used below.

\section{Extremal barrier and the two gaps}
The CA and CV calculations establish the structure of the superextremal side. We now connect this to extremality. The CA complexity of formation of charged AdS black holes diverges
logarithmically in the extremal limit~\cite{Carmi2017}. The same logarithmic divergence appears in CA
and CV for rotating black holes approaching extremality~\cite{Balushi2020,Bernamonti2021}.
Consequently, prescription by prescription, $\Cf(\Ext)$ is divergent while $\Cf(\Bsbx)$ is finite for
every fixed subextremal member. Applying Eq.~\eqref{eq:reverse-triangle} to the common vacuum reference
gives
\begin{equation}
\Crel(\Bsbx,\Ext)\geq
\left|\Cf(\Ext)-\Cf(\Bsbx)\right|\to\infty .
\label{eq:first-gap}
\end{equation}
This is a holographic complexity analogue of the third-law of black hole mechanics along the stationary family:
under finite complexity growth, the extremal state cannot be prepared from a fixed
subextremal member in finite boundary time. Accordingly, any genuinely dynamical formation mechanism, including those considered in~\cite{KehleUnger2025,Gadioux2026}, must violate at least one assumption of the argument, for example, the assumption of finite complexity growth.

For WCCC, the relevant channel is the black-hole-mediated path in which one attempts to move from
$\Bsbx$ across the extremality bound into $\Bsprx$, as in Wald-type overcharging experiments
\cite{Wald1974,Hubeny1999,SorceWald2017}. Such a continuous and quasi-stationary path must encounter the stationary
extremal surface separating the black-hole and superextremal sectors. The first gap already obstructs
the approach to $\Ext$ under finite complexity growth. The explicit calculation supplies the second
gap independently. In minimal CA, $\Cf^{\rm A}(\Bsprx)$ is finite after the same UV subtraction as
for the adjacent states; in CV, $\Cf^{\rm V}(\Bsprx)$ is also finite. Therefore the following bound
applies separately with $\Cf=\Cf^{\rm A}$ or $\Cf=\Cf^{\rm V}$:
\begin{align}
\Crel(\Ext,\Bsprx)
&\geq\left|\Cf(\Ext)-\Cf(\Bsprx)\right|\notag\\
&\to\infty .
\label{eq:second-gap}
\end{align}
For $\gamma\ne0$, the raw CA action instead contains the prescription-dependent local power law of
Eq.~\eqref{eq:fulldivcombined}. Since this term is not the universal extremal-throat logarithm, it cannot provide a scheme-independent cancellation of that logarithm. Instead, the additional power-law divergence further enlarges the complexity separation between the extremal and superextremal configurations. Thus, within this prescription, the complexity-based obstruction to crossing extremality is strengthened rather than weakened.

This complexity obstruction is complementary to, not a replacement for, the classical overcharging and
overspinning literature. Apparent test-particle overcharging windows motivate the problem
\cite{Hubeny1999}, whereas absorption constraints, test-field analyses, and higher-order backreaction
protect broad classes of examples~\cite{Wald1974,Natario2016,Semiz2011,Gwak2018,SorceWald2017}. The present point is different: in AdS holography, the stationary parameter space itself exhibits a complexity-geometric barrier at extremality. The result therefore applies directly to the black-hole-mediated quasistationary route, which passes through the extremal configuration. A finite relative complexity between a subextremal black hole and a superextremal configuration would not by itself establish that a corresponding physical transition exists, since relative complexity characterizes computational accessibility rather than dynamical admissibility. Conversely, the divergence found at extremality obstructs any preparation protocol for which this relative complexity provides an operational lower bound and the complexity growth rate remains finite. Boundary-sourced AdS singularities such as those studied in~\cite{Horowitz2016}, pinch-off singularities, and far-from-stationary processes lie outside what is
proven here.

\section{Universality}
The above results are not specific to overcharged RN--AdS, but follow more generally. For any
static, spherically symmetric geometry of the form in Eq.~\eqref{eq:metric} with
\begin{equation}
f(r)\sim a r^{-p},\qquad a>0,
\label{eq:scaling}
\end{equation}
near $r=0$, the WdW time width remains finite for $p>-1$, while Eq.~\eqref{eq:ghy-combined}
gives
\begin{equation}
I_{\rm GHY}(\epsilon)\sim \frac{\Om\Delta t_0}{8\pi\GN}a\left(\frac{p}{2}-(D-2)\right)
\epsilon^{D-3-p}.
\label{eq:ghy-scaling}
\end{equation}
Hence, for $p>D-3$ the GHY term diverges as a power law; in the marginal case $p=D-3$ it remains
finite, consistent with negative-mass Schwarzschild--AdS~\cite{Katoch2023}; and for $p<D-3$ it
vanishes. For charged solutions the Maxwell and on-shell Einstein--Hilbert bulk terms contribute
additional power-law terms of the same order, as shown above. For genuine Einstein--Maxwell solutions
the total may cancel as it does for RN--AdS in the minimal prescription; whether that cancellation
persists in a given theory must be checked from its on-shell structure.

The statement that is robust across this static scaling class is weaker but sufficient: repeating the
explicit endpoint evaluation above with $f\sim ar^{-p}$ makes the joint--counterterm sum proportional
to $\epsilon^{D-2}\ln\epsilon$, while every remaining near-singularity contribution is a pure power of
$\epsilon$. No near-origin term generates an unsuppressed logarithm. Thus the local singularity
structure cannot supply a universal counterpart of the extremal throat logarithm. This applies to four-dimensional Reissner--Nordstr\"om black holes
($p=2$), their higher-dimensional charged generalizations ($p=2D-6$), and broader classes of
multi-charge or dilatonic charged solutions whenever the same near-singularity scaling holds. Theories with
dilatonic or axionic couplings that modify the near-origin scaling should be checked case by case.

In CV, for $f(r)\sim ar^{-p}$, the near-singularity contribution is
\begin{equation}
V(0,\epsilon)\sim \frac{\Om}{\sqrt a}
\int_0^\epsilon dr\,r^{D-2+p/2}
=O(\epsilon^{D-1+p/2}).
\label{eq:cv-universality}
\end{equation} Thus, no logarithmic CV divergence appears whenever $D-1+p/2>0$, including
RN--AdS with $p=2D-6$.

\section{Conclusion and discussion}
The analysis identifies two successive divergences on relative complexity that provide a holographic
complexity-geometric obstruction to crossing the stationary extremal surface. The first follows from
finite subextremal complexity of formation and the logarithmically divergent extremal value. The second
follows, in CA and independently in CV, because the corresponding superextremal complexity of
formation is either finite or power-law divergent, and therefore cannot remove the logarithmic divergence of the extremality. Framing the result through
Eq.~\eqref{eq:reverse-triangle} avoids equating a difference of formation complexities with the exact
optimal transition circuit: the difference is used only as a rigorous lower bound on relative
complexity~\cite{BrownSusskind2018,Nielsen2006,Auzzi2021}.

The near-singularity action calculation is fully local. The Maxwell bulk, on-shell
Einstein--Hilbert, and GHY terms contribute $+(D-2)$, $-(D-4)$, and $-2$ in units of $u$, and cancel in
the minimal prescription. The two null--timelike joints and both null counterterms have been integrated
explicitly; their affine-normalization dependence cancels and their combined contribution vanishes as
$\epsilon^{D-2}\ln\epsilon$. The Maxwell boundary term is the only source of prescription dependence
at leading order and produces a regulator-local power law rather than a logarithm.

Under the stated finite-growth assumption, no quasi-stationary boundary evolution can traverse
$\Bsbx\to\Ext\to\Bsprx$ in finite time within CA or CV. This is not a proof of full WCCC and is
not a universal classical third law. It is a conditional statement about the stationary holographic
family, compatible with known dynamical constructions of extremal horizons
\cite{KehleUnger2025,Gadioux2026} and with conditional AdS third-law results
\cite{McSharryReall2025}. The sourced, far-from-equilibrium, and topology-changing channels remain
outside its scope.

The finite-growth assumption is essential and also limits any extension to de Sitter holography. In
de Sitter static-patch implementations, CA and CV can display hyperfast behavior, with the complexity and
its growth rate diverging at a finite critical time~\cite{Jorstad2022}. In such a prescription the inference
from divergent relative complexity to infinite preparation time does not follow, because the assumed finite
growth rate is violated. Other generalized de Sitter complexity functionals exhibit different late-time
behavior~\cite{AguilarGutierrez2024}, emphasizing the prescription dependence. Moreover, the AdS vacuum
subtraction and the standard unitary boundary-time interpretation used in the present argument are not
available in the same form in asymptotically de Sitter settings. We therefore make no de Sitter claim.

Finally, CA and CV are conjectural and are not unique among bulk observables with complexity-like
properties~\cite{Belin2022}. Agreement of the logarithmic mismatch in both prescriptions makes the
present mechanism less dependent on either one individually, but a direct derivation of the barrier
from the dual quantum theory remains the decisive open problem. Rotating superextremal geometries are
another natural test, although the Kerr--AdS ring singularity requires a separate treatment of the WdW
action.

\begin{acknowledgments}
It is a pleasure to thank Chris Akers, Koray D\"uzta\c{s}, \.Ibrahim Semiz, Asl\i{} Tuncer, Nicol\`{o} Zenoni, and Ying Zhao for
useful discussions and correspondence.
\end{acknowledgments}

\appendix
\section{Conventions and the exact cancellation}
\label{app:total-derivative}
With the action normalization in Eq.~\eqref{eq:bulk-action}, the Einstein equation is
\begin{equation}
R_{\mu\nu}-\frac{1}{2}Rg_{\mu\nu}+\Lambda g_{\mu\nu}
=2F_{\mu\alpha}F_\nu{}^\alpha-\frac{1}{2}g_{\mu\nu}F^2.
\end{equation}
Taking the trace gives
\begin{equation}
R=\frac{2D\Lambda}{D-2}+\frac{D-4}{D-2}F^2,
\end{equation}
and hence Eq.~\eqref{eq:trace}. The Maxwell equation for Eq.~\eqref{eq:gaugefield} is
$\partial_r(r^{D-2}F^{rt})=0$, and the $tt$ Einstein equation fixes
$c_D^2=(D-2)(D-3)/2$, giving Eq.~\eqref{eq:f}.

The cancellation in Eq.~\eqref{eq:exact-cancel} can be seen directly. Per unit coordinate time and unit
sphere volume, the on-shell bulk integrand is
\begin{equation}
\begin{split}
&\frac{r^{D-2}}{16\pi\GN}\left(R-2\Lambda-F^2\right)\\
&\quad=\frac{1}{16\pi\GN}\frac{d}{dr}
\left[\frac{4\Lambda r^{D-1}}{(D-1)(D-2)}-\frac{2q^2}{r^{D-3}}\right].
\end{split}
\label{eq:total-derivative}
\end{equation}
After integration over the WdW patch, $\Delta t(r)=\Delta t_0+O(r^{2D-5})$, so the divergent lower-endpoint contribution from Eq.~\eqref{eq:total-derivative} is
$+2q^2\Delta t_0/(16\pi\GN\epsilon^{D-3})$ per unit sphere volume, namely $+2u$. The GHY bracket in Eq.~\eqref{eq:ghy-expanded} contributes $-q^2/\epsilon^{D-3}$ at the same surface, namely $-2u$. The two terms are equal and opposite for every $D\ge4$. The subleading terms, including the cosmological contribution, the $\mu$-dependent term, and the corrections from $\Delta t(r)-\Delta t_0$, are finite or vanish at the regulator and do not affect the divergence analysis.

The Maxwell boundary term gives the only prescription-dependent near-origin divergence. On the regulator surface, $d\Sigma_r=-\epsilon^{D-2}dt\,d\Omega_{D-2}$ with the same outward normal convention, and
\begin{equation}
F^{rt}A_t=F_{tr}A_t=-A_t\partial_r A_t=-\frac{1}{2}\partial_r A_t^2.
\end{equation}
Its divergent part is fixed by the field strength,
\begin{equation}
F_{tr}A_t=\frac{(D-2)q^2}{2r^{2D-5}}-c_D\Phi\frac{q}{r^{D-2}},
\end{equation}
so the regulator contribution is
\begin{equation}
I_\gamma(r=\epsilon)=-\frac{\gamma\Om\Delta t_0}{4\pi\GN}
\left[\frac{(D-2)q^2}{2\epsilon^{D-3}}-c_D\Phi q+O(\epsilon)\right],
\end{equation}
which is Eq.~\eqref{eq:gamma-div}. A gauge shift changes only the finite part. The null segments of
$I_\gamma$ are finite near $r=0$, so no Maxwell boundary prescription generates a logarithmic divergence.

\bibliographystyle{apsrev4-2}

\begin{thebibliography}{99}
\bibitem{Penrose1969}
R.~Penrose,
\emph{Gravitational Collapse: The Role of General Relativity},
Riv.\ Nuovo Cim.\ \textbf{1}, 252--276 (1969).

\bibitem{Maldacena1998}
J.~M.~Maldacena,
\emph{The Large $N$ Limit of Superconformal Field Theories and Supergravity},
Adv.\ Theor.\ Math.\ Phys.\ \textbf{2}, 231--252 (1998).

\bibitem{Brown2016PRL}
A.~R.~Brown, D.~A.~Roberts, L.~Susskind, B.~Swingle, and Y.~Zhao,
\emph{Holographic Complexity Equals Bulk Action?},
Phys.\ Rev.\ Lett.\ \textbf{116}, 191301 (2016).

\bibitem{Brown2016PRD}
A.~R.~Brown, D.~A.~Roberts, L.~Susskind, B.~Swingle, and Y.~Zhao,
\emph{Complexity, Action, and Black Holes},
Phys.\ Rev.\ D \textbf{93}, 086006 (2016).

\bibitem{SusskindStanford}
D.~Stanford and L.~Susskind,
\emph{Complexity and Shock Wave Geometries},
Phys.\ Rev.\ D \textbf{90}, 126007 (2014).

\bibitem{ChapmanFormation2017}
S.~Chapman, H.~Marrochio, and R.~C.~Myers,
\emph{Complexity of Formation in Holography},
JHEP \textbf{01}, 062 (2017).

\bibitem{ChapmanPolicastro2022}
S.~Chapman and G.~Policastro,
\emph{Quantum Computational Complexity from Quantum Information to Black Holes and Back},
Eur.\ Phys.\ J.\ C \textbf{82}, 128 (2022).

\bibitem{BrownSusskind2018}
A.~R.~Brown and L.~Susskind,
\emph{The Second Law of Quantum Complexity},
Phys.\ Rev.\ D \textbf{97}, 086015 (2018).

\bibitem{Nielsen2006}
M.~A.~Nielsen, M.~R.~Dowling, M.~Gu, and A.~C.~Doherty,
\emph{Quantum Computation as Geometry},
Science \textbf{311}, 1133--1135 (2006).

\bibitem{Auzzi2021}
R.~Auzzi, S.~Baiguera, G.~B.~De Luca, A.~Legramandi, G.~Nardelli, and N.~Zenoni,
\emph{Geometry of Quantum Complexity},
Phys.\ Rev.\ D \textbf{103}, 106021 (2021).

\bibitem{Haferkamp2021}
J.~Haferkamp, P.~Faist, N.~B.~T.~Kothakonda, J.~Eisert, and N.~Yunger Halpern,
\emph{Linear Growth of Quantum Circuit Complexity},
Nat.\ Phys.\ \textbf{18}, 528--532 (2022).

\bibitem{Brandao2021}
F.~G.~S.~L.~Brand\~ao, W.~Chemissany, N.~Hunter-Jones, R.~Kueng, and J.~Preskill,
\emph{Models of Quantum Complexity Growth},
PRX Quantum \textbf{2}, 030316 (2021).

\bibitem{Lloyd2000}
S.~Lloyd,
\emph{Ultimate Physical Limits to Computation},
Nature \textbf{406}, 1047--1054 (2000).

\bibitem{Engelhardt2025}
N.~Engelhardt, \AA.~K.~Folkestad, A.~Levine, E.~Verheijden, and L.~Yang,
\emph{Cryptographic Censorship},
JHEP \textbf{01}, 122 (2025).

\bibitem{Alishahiha2022}
M.~Alishahiha, S.~Banerjee, J.~Kames-King, and E.~Loos,
\emph{Complexity as a Holographic Probe of Strong Cosmic Censorship},
Phys.\ Rev.\ D \textbf{105}, 026001 (2022).

\bibitem{Mohan2025}
V.~Mohan,
\emph{Black Hole Singularities from Holographic Complexity},
JHEP \textbf{07}, 275 (2025).

\bibitem{Akers2024}
C.~Akers, N.~Engelhardt, D.~Harlow, G.~Penington, and S.~Vardhan,
\emph{The Black Hole Interior from Non-Isometric Codes and Complexity},
JHEP \textbf{06}, 155 (2024).

\bibitem{Bardeen1973}
J.~M.~Bardeen, B.~Carter, and S.~W.~Hawking,
\emph{The Four Laws of Black Hole Mechanics},
Commun.\ Math.\ Phys.\ \textbf{31}, 161--170 (1973).

\bibitem{Israel1986}
W.~Israel,
\emph{Third Law of Black-Hole Dynamics: A Formulation and Proof},
Phys.\ Rev.\ Lett.\ \textbf{57}, 397--399 (1986).

\bibitem{KehleUnger2025}
C.~Kehle and R.~Unger,
\emph{Gravitational Collapse to Extremal Black Holes and the Third Law of Black Hole Thermodynamics},
J.\ Eur.\ Math.\ Soc., published online 27 January 2025.

\bibitem{Gadioux2026}
M.~Gadioux, H.~S.~Reall, and J.~E.~Santos,
\emph{Formation of Extremal Reissner--Nordstr\"om Black Holes: Insights from Numerics},
Phys.\ Rev.\ D \textbf{113}, 084001 (2026).

\bibitem{McSharryReall2025}
A.~M.~McSharry and H.~S.~Reall,
\emph{Supersymmetric Black Holes and the Third Law of Black Hole Mechanics},
Phys.\ Rev.\ D \textbf{112}, 104009 (2025).

\bibitem{Carmi2017}
D.~Carmi, S.~Chapman, H.~Marrochio, R.~C.~Myers, and S.~Sugishita,
\emph{On the Time Dependence of Holographic Complexity},
JHEP \textbf{11}, 188 (2017).

\bibitem{Balushi2020}
A.~Al Balushi, R.~A.~Hennigar, H.~K.~Kunduri, and R.~B.~Mann,
\emph{Holographic Complexity of Rotating Black Holes},
JHEP \textbf{05}, 226 (2021).

\bibitem{Bernamonti2021}
A.~Bernamonti, F.~Bigazzi, D.~Billo, L.~Faggi, and F.~Galli,
\emph{Holographic and QFT Complexity with Angular Momentum},
JHEP \textbf{11}, 037 (2021).

\bibitem{Goto2019}
K.~Goto, H.~Marrochio, R.~C.~Myers, L.~Queimada, and B.~Yoshida,
\emph{Holographic Complexity Equals Which Action?},
JHEP \textbf{02}, 160 (2019).

\bibitem{Lehner2016}
L.~Lehner, R.~C.~Myers, E.~Poisson, and R.~D.~Sorkin,
\emph{Gravitational Action with Null Boundaries},
Phys.\ Rev.\ D \textbf{94}, 084046 (2016).

\bibitem{ReynoldsRoss2017}
A.~Reynolds and S.~F.~Ross,
\emph{Divergences in Holographic Complexity},
Class.\ Quantum Grav.\ \textbf{34}, 105004 (2017).

\bibitem{Katoch2023}
G.~Katoch, J.~Ren, and S.~R.~Roy,
\emph{Quantum Complexity and Bulk Timelike Singularities},
JHEP \textbf{12}, 085 (2023).

\bibitem{Wald1974}
R.~M.~Wald,
\emph{Gedanken Experiments to Destroy a Black Hole},
Ann.\ Phys.\ \textbf{82}, 548--556 (1974).

\bibitem{Hubeny1999}
V.~E.~Hubeny,
\emph{Overcharging a Black Hole and Cosmic Censorship},
Phys.\ Rev.\ D \textbf{59}, 064013 (1999).

\bibitem{SorceWald2017}
J.~Sorce and R.~M.~Wald,
\emph{Gedanken Experiments to Destroy a Black Hole. II. Kerr--Newman Black Holes Cannot Be Over-Charged or Over-Spun},
Phys.\ Rev.\ D \textbf{96}, 104014 (2017).

\bibitem{Natario2016}
J.~Nat\'ario, L.~Queimada, and R.~Vicente,
\emph{Test Fields Cannot Destroy Extremal Black Holes},
Class.\ Quantum Grav.\ \textbf{33}, 175002 (2016).

\bibitem{Semiz2011}
\.I.~Semiz,
\emph{Dyonic Kerr--Newman Black Holes, Complex Scalar Field and Cosmic Censorship},
Gen.\ Relativ.\ Gravit.\ \textbf{43}, 833--846 (2011).

\bibitem{Gwak2018}
B.~Gwak,
\emph{Weak Cosmic Censorship Conjecture in Kerr--(Anti-)de Sitter Black Hole with Scalar Field},
JHEP \textbf{09}, 081 (2018).

\bibitem{Horowitz2016}
G.~T.~Horowitz, J.~E.~Santos, and B.~Way,
\emph{Evidence for an Electrifying Violation of Cosmic Censorship},
Class.\ Quantum Grav.\ \textbf{33}, 195007 (2016).

\bibitem{Jorstad2022}
E.~J{\o}rstad, R.~C.~Myers, and S.-M.~Ruan,
\emph{Holographic Complexity in $\mathrm{dS}_{d+1}$},
JHEP \textbf{05}, 119 (2022).

\bibitem{AguilarGutierrez2024}
S.~E.~Aguilar-Gutierrez, M.~P.~Heller, and S.~Van der Schueren,
\emph{Complexity Equals Anything Can Grow Forever in de Sitter Space},
Phys.\ Rev.\ D \textbf{110}, 066009 (2024).

\bibitem{Belin2022}
A.~Belin, R.~C.~Myers, S.-M.~Ruan, G.~S\'arosi, and A.~J.~Speranza,
\emph{Does Complexity Equal Anything?},
Phys.\ Rev.\ Lett.\ \textbf{128}, 081602 (2022).

\end{thebibliography}

\end{document}